\documentclass[traditabstract]{aa}

\usepackage{amsmath}
\usepackage{amssymb}
\usepackage{txfonts}
\usepackage{epsfig,graphicx}
\usepackage[colorlinks=true, linkcolor=blue, citecolor=blue, urlcolor=blue]{hyperref}
\usepackage{natbib}
\bibpunct{(}{)}{;}{a}{}{,} 
\usepackage{color}

\begin{document}

\title{The bolometric and UV attenuation in normal spiral galaxies of the \textit{Herschel}\thanks{\textit{Herschel} is an ESA space observatory with science instruments provided by European-led Principal Investigator consortia and with important participation from NASA.} Reference Survey}

\titlerunning{The attenuation by dust in spiral galaxies}

\author{
S. Viaene\inst{\ref{inst-UGent}}
\and
M. Baes\inst{\ref{inst-UGent}} 
\and
G. Bendo\inst{\ref{inst-Manchester}}
\and
M. Boquien\inst{\ref{inst-Chile}, \ref{inst-Cambridge}}
\and
A. Boselli\inst{\ref{inst-Marseille}}
\and 
L. Ciesla\inst{\ref{inst-Saclay}}
\and
L. Cortese\inst{\ref{inst-UWA}} 
\and
I. De Looze\inst{\ref{inst-UGent},\ref{inst-Cambridge},\ref{inst-London}}
\and
S. Eales\inst{\ref{inst-Cardiff}}
\and
J. Fritz\inst{\ref{inst-UGent},\ref{inst-Morelia}} 
\and
O. \L. Karczewski\inst{\ref{inst-Saclay}}
\and
S. Madden\inst{\ref{inst-Saclay}}
\and
M. W. L. Smith\inst{\ref{inst-Cardiff}}
\and
L. Spinoglio\inst{\ref{inst-Rome}}
}

\institute{
Sterrenkundig Observatorium, Universiteit Gent, Krijgslaan 281, B-9000 Gent, Belgium\label{inst-UGent} \\
\email{sebastien.viaene@ugent.be}
\and
UK ALMA Regional Centre Node, Jodrell Bank Centre for Astrophysics, School of Physics and Astronomy, University of Manchester, Oxford Road, Manchester M13 9PL, UK\label{inst-Manchester}
\and
Unidad de Astronomía, Fac. de Ciencias Básicas, Universidad de Antofagasta, Antofagasta, Chile\label{inst-Chile}
\and 
Institute of Astronomy, University of Cambridge, Madingley Road, Cambridge, CB3 0HA, UK\label{inst-Cambridge}
\and
Aix Marseille Universit\'e, CNRS, LAM (Laboratoire d'Astrophysique de Marseille) UMR 7326, 13388 Marseille, France\label{inst-Marseille} 
\and
Laboratoire AIM, CEA/DSM--CNRS--Universit\'{e} Paris Diderot, IRFU/Service d'Astrophysique, CEA Saclay, 91191 Gif-sur-Yvette, France\label{inst-Saclay}
\and
International Centre for Radio Astronomy Research, University of Western Australia, 35 Stirling Highway, Crawley, WA 6009, Australia\label{inst-UWA}
\and
Department of Physics and Astronomy, University College London, Gower Street, London WC1E 6BT, UK\label{inst-London}
\and
School of Physics \& Astronomy, Cardiff University, The Parade, Cardiff, CF24 3AA, UK\label{inst-Cardiff}
\and
Instituto de Radioastronom\'{\i}a y Astrof\'{\i}sica, UNAM, Antigua Carretera a P\'atzcuaro \# 8701, Morelia, Michoac\'an, Mexico\label{inst-Morelia}
\and
Istituto di Fisica dello Spazio Interplanetario, INAF, Via del Fosso del Cavaliere 100, I-00133 Roma, Italy\label{inst-Rome}
}

\abstract{The dust in nearby galaxies absorbs a fraction of the UV-optical-near-infrared radiation produced by stars. This energy is consequently re-emitted in the infrared. We investigate the portion of the stellar radiation absorbed by spiral galaxies from the HRS by modelling their UV-to-submillimetre spectral energy distributions. Our models provide an attenuated and intrinsic SED, from which we find that on average 32 $\%$ of all starlight is absorbed by dust. We define the UV heating fraction as the percentage of dust luminosity that comes from absorbed UV photons and find this to be 56 $\%$, on average. This percentage varies with morphological type, with later types having significantly higher UV heating fractions. We find a strong correlation between the UV heating fraction and specific star formation rate and provide a power-law fit. Our models allow us to revisit the IRX - $A_{FUV}$ relations, and derive these quantities directly within a self-consistent framework. We calibrate this relation for different bins of $NUV-r$ colour and provide simple relations to relate these parameters. We investigated the robustness of our method and conclude that the derived parameters are reliable within the uncertainties that are inherent to the adopted SED model. This calls for a deeper investigation of how well extinction and attenuation can be determined through panchromatic SED modelling.}


\maketitle
\section{Introduction}

The obscuring power of cosmic dust has a significant effect on our view of the Universe. Dust grains in all environments absorb a portion of the light emitted by stars and active galactic nuclei (AGN). This energy is reprocessed and makes dust the prime source of emission in the far-IR (FIR) and sub-millimetre (submm) regimes. 
For a given galaxy, it is hard to measure exactly how much of the starlight is attenuated by dust and how this amount differs with wavelength and environment. These are useful quantities to know, for example when constructing luminosity functions and colour-magnitude diagrams and when determining star formation rates (SFRs) or stellar masses. Additionally, it can play a key role in comparing simulated galaxies or universes with our own Universe. In a broader context, the average fraction of absorbed energy (and the corresponding dust luminosity) allows estimates of the extragalactic background light \citep[see e.g.][]{Skibba2011}.

Several studies in the 1990s attempted to determine the bolometric luminosity in nearby galaxies based on data from the IRAS mission, and they found a typical number of about 30\% for the fraction of the bolometric luminosity absorbed and re-emitted by dust \citep{1991AJ....101..354S, 1995A&A...293L..65X}. A dedicated study was performed by \citet{2002MNRAS.335L..41P}. Their work was based on a sample of 28 spiral galaxies with multi-band optical and ISOPHOT FIR observations. They found the same percentage of attenuation: on average, interstellar dust absorbs and re-emits about 30\% of the bolometric luminosity of late-type galaxies. 

While the study of \citet{2002MNRAS.335L..41P} was a significant step forward in determining the bolometric attenuation compared to IRAS-based estimates, their work still suffered from a number of limitations. Their sample size was modest, which limits a detailed statistical study. Moreover, the wavelength range  on which their analysis was based was rather limited. Optical data were available for all galaxies, but UV\footnote{Following \citealt{1995A&A...293L..65X}, we define the UV domain as the wavelength range up to $3650~\AA$} data were not. In the near-IR (NIR), only {\it{K}}'-band magnitudes were available ({\it{H}}-band data for only a few galaxies). But more importantly, the dust luminosity was estimated using only three FIR bands centred at 60, 100, and 170 $\mu$m. The lack of data beyond 170~$\mu$m and in the mid-IR (MIR) region puts limitations on a secure determination of the dust luminosity. Finally, only Virgo Cluster galaxies were included in the study, which may cause significant bias (e.g. due to the dense cluster environment). 

More than a decade later, we are now in a situation where we can eliminate most of these concerns. Most importantly, the number of galaxies for which the entire UV-submm spectral energy distribution has been sampled accurately and densely, has increased enormously in the past few years. Concerning the stellar part of the SEDs, thousands of nearby galaxies have been imaged in the UV by the Galaxy Evolution Exporer \citep[\textit{GALEX},][]{2005ApJ...619L...1M}, in the optical by the Sloan Digital Sky Survey \citep[SDSS,][]{2011ApJS..193...29A}, and in the NIR by the Two-Micron All Sky Survey \citep[2MASS,][]{2006AJ....131.1163S}. Even more important for our goal is the spectacular increase in the coverage of the infrared part of the SED. MIR imaging is now available for the entire sky thanks to the WISE survey \citep{2010AJ....140.1868W}, and many nearby galaxies have been imaged by \textit{Spitzer} \citep{2004ApJS..154....1W} and \textit{Herschel} \citep{2010A&A...518L...1P} at longer wavelengths. The latter mission is particularly useful for our goals as \textit{Herschel} is the first large-scale mission to cover the submm region at wavelengths beyond 200~$\mu$m, where the emission from cool dust dominates.

Consequently, the exercise of determining the bolometric absorption by dust was repeated in the \textit{Herschel} era. \citet{Skibba2011} investigated the dust/stellar flux ratio for the KINGFISH sample \citep{2011PASP..123.1347K}. Only considering the spiral galaxies of their sample, they find an equivalent of   $31 \%$ for the bolometric attenuation. Although this is a comforting confirmation of previous studies, it was still derived in a rather empirical way from (as they state) a relatively small and incomplete sample.

In addition to the availability of large samples of galaxies with high-quality SED data, we now also have the advantage that advanced tools have become available for analysing observed SEDs. Panchromatic SED fitting has been the subject of intense research in the past few years \citep[for an overview, see][]{2011Ap&SS.331....1W, 2013ARA&A..51..393C}. Various powerful libraries and fitting tools have become available to analyse the observed SEDs from UV to submm wavelengths \citep[e.g.][Boquien et al. in prep]{2008ApJS..176..438G, daCunha2008, Noll2009, 2011ApJ...740...22S, 2011MNRAS.410.2043S}.

The combination of high-quality SED data and advanced panchromatic SED fitting tools implies that we can extend our analysis beyond only determining the bolometric attenuation in galaxies. First, we can correlate the bolometric attenuation with fundamental physical galaxy properties derived from the SED modelling, such as stellar masses and SFRs. Second, it also allows a more detailed investigation of the absorbed energy spectrum. This can be used to quantify the importance of different heating sources for the dust in normal spiral galaxies. This question has been the subject of quite some debate in the recent past, and different approaches have been used to tackle this problem \citep[e.g.][]{2007ApJ...663..866D, 2010MNRAS.409....2R,2010A&A...518L..61B, 2012A&A...540A..54B, 2010A&A...518L..65B, 2012MNRAS.419.1833B, 2015MNRAS.448..135B, 2011AJ....142..111B, 2011A&A...527A.109P, 2012MNRAS.427.2797D, 2012MNRAS.419..895D, 2013A&A...549A..70H}. The results from these studies are mixed: it seems that in normal spiral galaxies, both young and evolved stellar populations can dominate the dust heating budget. 

Another important attenuation-related quantity is the attenuation in the \textit{GALEX} $FUV$ band,  $A_{FUV}$. Estimating the attenuation in this wavelength range is extremely useful if one wants to determine the SFR of galaxies, often probed by the $FUV$ flux. Closely related to $A_{FUV}$ is the ratio of the total IR (TIR) and the $FUV$ luminosity; IRX. The IRX ratio is one of the best, geometry-independent methods to determine the UV attenuation in galaxies \citep[see e.g.][and references therein]{Buat1996, Meurer1999,Gordon2000}. Recent studies have attempted to calibrate this relation based on theoretical models and observations \citep{Cortese2008,Hao2011}. \citet{Boquien2012} investigated the IRX-$\beta$ relation at sub-galactic scales for a sample of seven face-on galaxies. They made use of the powerful tool of panchromatic SED fitting to provide a consistent picture of the attenuation and its link to observational quantities. A calibration of $A_{FUV}$ for a large and representative sample of local galaxies using this technique would prove its worth. With our panchromatic SED modelling and information on the intrinsic flux of each galaxy, we approach the independent measurement of $A_{FUV}$. This allows us to revisit previous calibrations and provide a new calibration based on physically consistent parameters for a representative set of local galaxies.

This paper aims to provide a better determination of three main attenuation-related galaxy parameters and to understand how these parameters are related to global galaxy properties. This is done in a self-consistent model framework and through panchromatic SED fitting. Our goal is to quantify how much of the starlight is absorbed by dust, both in total energy and in the UV domain, where the effect of extinction and dust heating is the greatest. In Section~\ref{sec:data} we present the sample we use for this study and the available data. In Section~\ref{sec:method} we discuss the methods we use for our analysis and define three main attenuation parameters: the bolometric dust fraction $f_\mathrm{dust}$, the UV heating fraction $\xi_\mathrm{UV}$, and the $FUV$ attenuation $A_{FUV}$. Our results are presented in Section~\ref{sec:results}. Section~\ref{sec:discussion} discusses the correlations found and highlights some of the model caveats. We conclude in Section~\ref{sec:conclusions}.

\section{Sample selection and data} \label{sec:data}

This paper is based on the \textit{Herschel} Reference Survey \citep[HRS,][]{2010PASP..122..261B}, a \textit{Herschel} guaranteed-time key programme that targeted a sample of 322 galaxies in the local Universe. The sample covers a wide range of densities from the field to the centre of the Virgo Cluster and spans the entire range of morphological types from ellipticals to late-type spirals. Thanks to the proximity (all galaxies are between 15 and 25 Mpc) and the completeness of the sample, the HRS can be used both for detailed studies of individual galaxies \citep[e.g.][]{2010A&A...518L..45G, 2010A&A...518L..63C, 2010A&A...518L..66R, 2010A&A...518L..72P, 2012MNRAS.427.2797D} and for statistical analyses \citep[e.g.][]{2010A&A...518L..61B, 2012A&A...540A..54B, 2012A&A...540A..52C, 2012ApJ...748..123S, Ciesla2014}. 

Thousands of nearby galaxies have been observed by \textit{Herschel} as part of several large surveys, and for many of these surveys complementary data are available from UV to MIR wavelengths. Our approach requires a representative sample of nearby galaxies with accurate photometry from the UV to submm wavelengths. We chose the HRS in particular is because of the $K$-band and volume-limited selection, making it a complete sample, representative of `normal' galaxies in the local universe.

In this paper we focus on the late-type galaxy subsample (types Sa and later, galaxy types are taken from Table~1 in \citet{2012A&A...544A.101C}). We exclude the galaxies without \textit{Herschel} PACS or SPIRE fluxes (HRS\,104, 116, 164, 195, 225, 228, 229, and 291) or without \textit{GALEX} fluxes (HRS\,4, 5, 10, 38, 73, 76, 238, 254, 259). Additionally, we remove HRS 284 (no SDSS) from our sample. Of the remaining galaxies, 14 were found to have a strong AGN. The AGN classification was done through analysis of optical spectral lines of the galaxy nuclei and will be presented in Gavazzi et al. (in prep.). We plot them separately in section \ref{sec:results}. However, we do not treat them as a separate subsample because, as we show, their attenuation properties are not fundamentally different. 

The final sample counts 239 galaxies. The spread over the different morphological types can be found in Table~{\ref{MeanValues.tab}}. The HRS was originally proposed to acquire \textit{Herschel} observations, but now a vast set of ancillary data is available. In this work, we focus on the broad-band data (see list below), but there is also narrow band photometry in H$\alpha$ \citep{Boselli2015}, optical spectroscopy \citep{2013A&A...550A.114B, 2013A&A...550A.115H}, and atomic and molecular gas data \citep{Boselli2014}. For the present study we use integrated flux densities over the UV-submm range:
\begin{itemize}
\item \textit{GALEX} $FUV$ and $NUV$ and SDSS {\em{gri}} flux densities are taken from \citet{2012A&A...544A.101C};
\item 2MASS {\em{J}}, {\em{H}}, and {\em{K}}$_{\text{s}}$ data from the 2MASS archive \citep{2MASS};
\item \textit{Spitzer} IRAC 3.6 and 4.5 $\mu$m flux densities from the S$^4$G project \citep{Sheth2010};
\item WISE 12 and 22 $\mu$m data taken from \citet{Ciesla2014};
\item Integrated \textit{Spitzer} MIPS 24, 70 and 160~$\mu$m fluxes are taken from \citet{2012MNRAS.423..197B};
\item \textit{Herschel} PACS 100 and 160 $\mu$m flux densities are taken from \citet{Cortese2014};
\item \textit{Herschel} SPIRE 250, 350, and 500 $\mu$m flux densities are taken from \citet{2012A&A...543A.161C}. Correction factors to convert them to the latest SPIRE calibration and beam factors are taken into account.
\end{itemize}
For many galaxies in the sample, additional data are available that could be added to the database. To have a database that is as uniform as possible for all galaxies in the sample, we limited our database to these data. 

\section{Dust attenuation analysis} \label{sec:method}

We make use of the Bayesian SED fitting code MAGPHYS \citep{daCunha2008}. The code has been used extensively in recent years and has proved reliable \citep[see e.g. ][and references therein]{Michalowski2014, Hayward2015,Smith2015}. The UV-to-submm SED is fitted by comparing a library of physically motivated SED templates to the observed data points. The parameter values used to build each template are then weighted by the corresponding $\chi^2$, creating probability density functions (PDF). As a result, we know the most probable value for each model parameter. The $\chi^2$ goodness of fit in MAGPHYS differs conceptually from a classic reduced $\chi^2$. The Bayesian approach of MAGPHYS does not actively vary a set of free parameters. Instead, a series of model templates are created a priori and then evaluated one-by-one with the observations. The weighted $\chi^2$ formalism described in \citet{daCunha2008} reflects the difference between model and observations and takes the observational uncertainty into account. It is set up in such a way that a $\chi^2$ around unity indicates a good fit.

The UV-to-NIR regime is modelled using the 2007 version of the \citet{Bruzual2003} simple stellar population formalism \citep[see ][]{Bruzual2007}. The model assumes a \citet{Chabrier2003} IMF and an exponentially declining star formation history (SFH), with random bursts of star formation superimposed. On top of that is the dust attenuation model based on the formalism of \cite{Charlot2000}. There is extinction and scattering by dust in the diffuse ISM and by dust in circumstellar clouds. The optical depth for a stellar population depends on its age, with young stars being more obscured than older stars. Galaxies with younger stellar populations will therefore have more extinction of their starlight (yielding higher dust luminosities) and are more likely to have higher values of attenuation. The MIR-to-submm SED is constructed using a series of modified black body functions and a template for the PAH features. The \citet{Charlot2000} dust model is used to treat the interaction of dust and starlight. A key point of MAGPHYS is the conservation of energy. The energy that is absorbed at UV and optical wavelengths is consistently re-emitted in the FIR and submm regimes. For more details on the theoretical model, we refer the reader to \citet{daCunha2008}.

We have adopted a slightly extended version of the code, which has a wider range in cold dust temperature ($T_C^\mathrm{ISM} = 10-30$K, compared to a standard range of $15-25$ K) and warm dust temperature ($T_W^\mathrm{BC} = 30-70$K, where the standard range is $30-60$ K). This extension is needed to adequately fit the SED objects for which the narrow standard dust temperature ranges are insufficient. For the HRS, cold dust temperatures were found to lie within the 10-30K range based on modified black body fitting \citep{Cortese2014}. The extended library of SED templates proved to work well at sub-kpc scales \citep{Viaene2014}, but also for integrated galaxy SEDs \citep[][Papallardo et al. in prep.]{Agius2015}.

\begin{figure}
\centering
\includegraphics[width=0.4\textwidth]{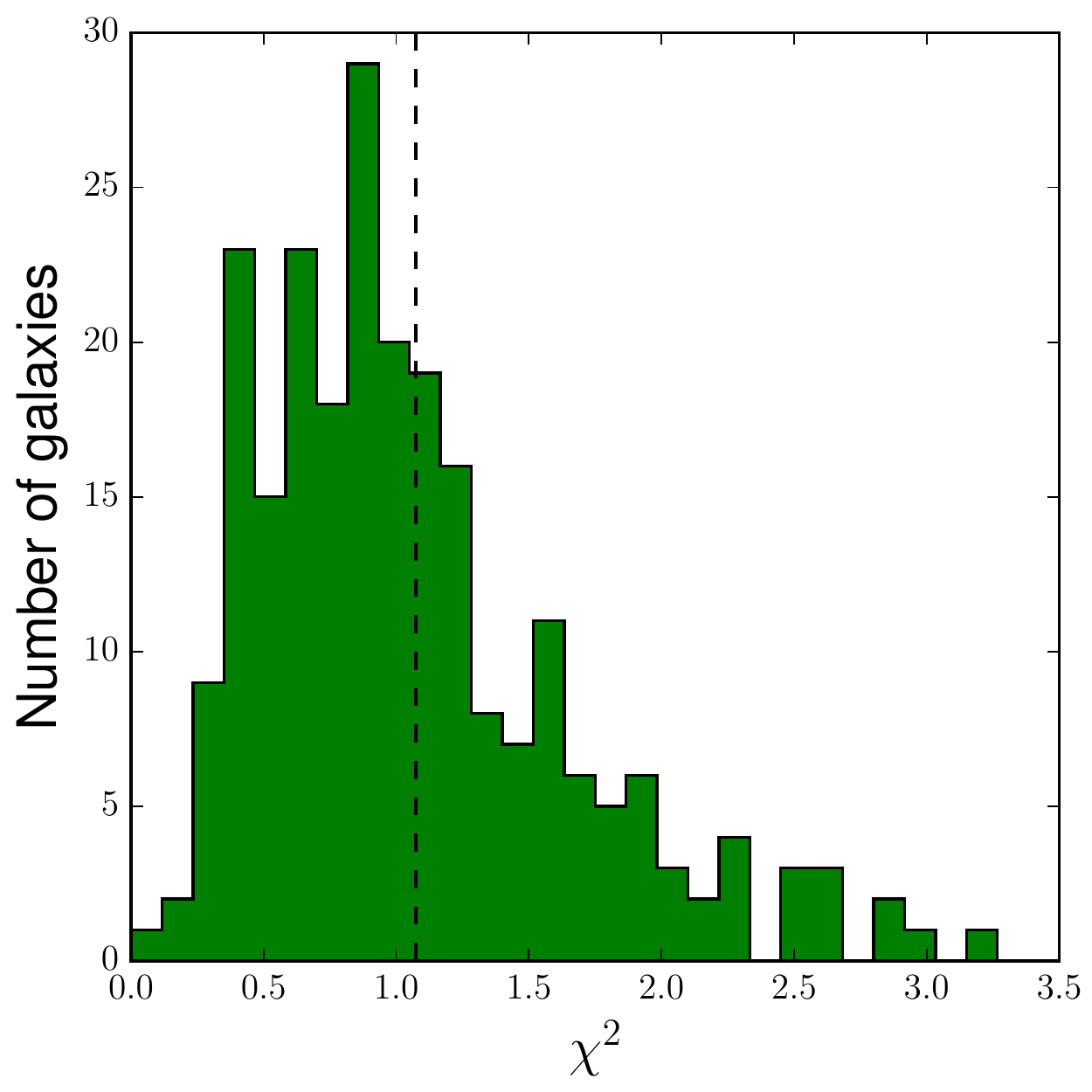}
\caption{$\chi^2$ distribution of the best fit SED models for our galaxy sample. The mean value of 1.07 is indicated by the dashed black line.}
\label{fig:hist_chi2}
\end{figure}

\begin{figure*}
\centering
\includegraphics[width=0.8\textwidth]{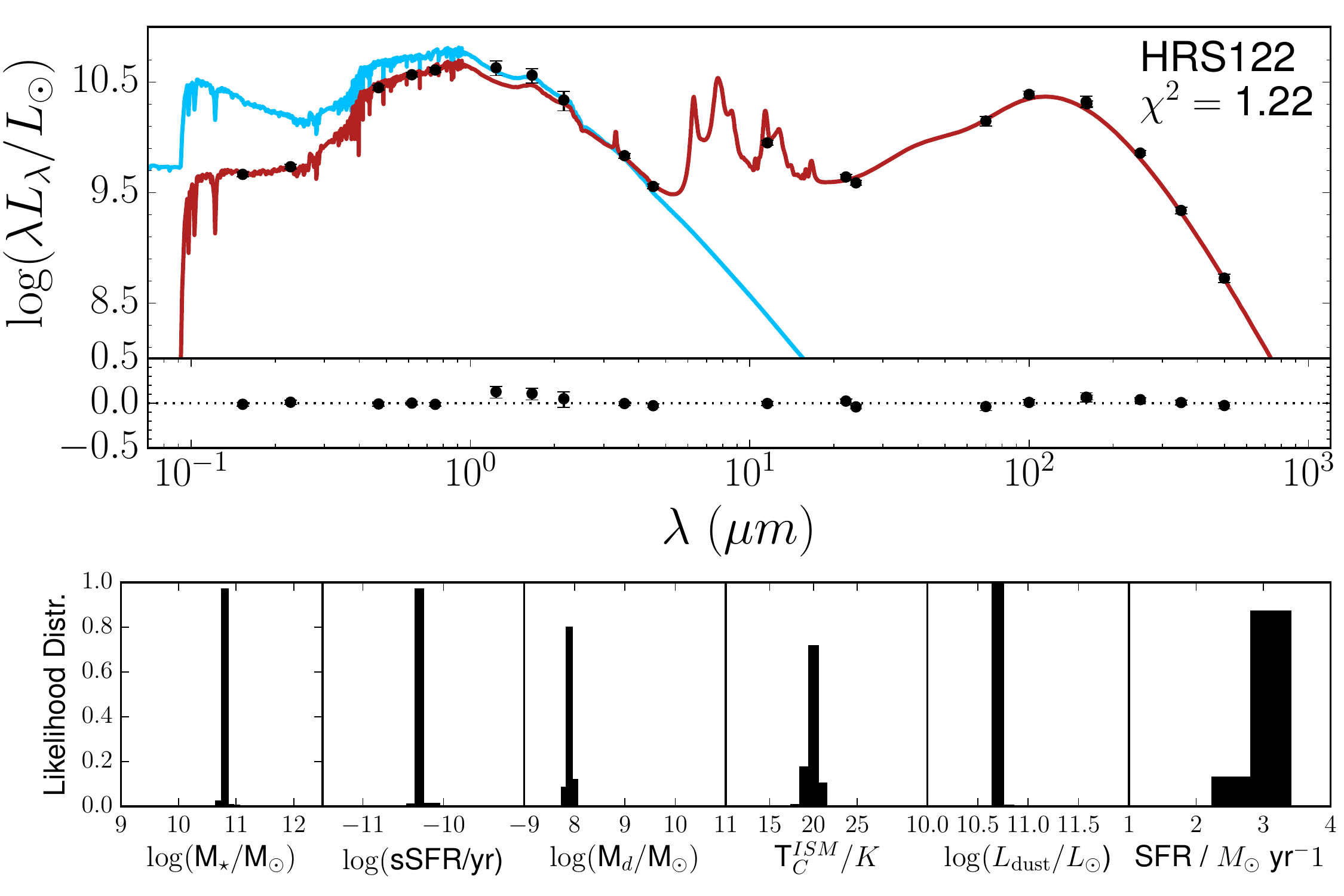}
\caption{SED and model of HRS 122 (M 100), which is representative for the sample. The top panel shows the data (black dots with error bars), the best fitting panchromatic SED fit (red line) and the intrinsic stellar SED (blue line). The residuals between data and best fitting model are shown below the SED. The bottom row shows the corresponding PDFs for several key parameters of the model.}
\label{MAGPHYSExamples.fig}
\end{figure*}

The quality of the fits to our 221 galaxies is generally excellent judging from a visual inspection of each galaxy and from the overall $\chi^2$ distribution (see Fig.~\ref{fig:hist_chi2}). It resembles a normal distribution, peaking close to unity, but has a small tail towards the high end. Two objects have a $\chi^2 > 3$: HRS 184 and HRS 222. Both objects show slightly offset MIPS data points, but the fits are still of sufficient quality for our purposes. We omit HRS 142 from our sample because MAGPHYS is unable to fit the exceptionally broad FIR peak ($\chi^2 > 18$). A representative SED fit is shown in Fig.~\ref{MAGPHYSExamples.fig}, together with several parameter PDFs.

MAGPHYS allows us to determine basic galaxy parameters, such as stellar mass ($M_\star$), SFR averaged over the last 100 Myr, and dust luminosity ($L_\mathrm{dust}$). Additionally, it provides us with the unattenuated SED. Upon comparing this intrinsic SED (without dust attenuation) with the `normal', attenuated SED, the wavelength dependence of dust attenuation can be investigated.

In this work, we make use of MAGPHYS' capacity to provide an attenuated and unattenuated SED. We focus on three attenuation-related quantities: 1) the bolometric attenuation $f_{\text{dust}}$ or the fraction of the bolometric luminosity that is absorbed by dust, 2) the UV heating fraction $\xi_{\text{UV}}$, which is the fraction of UV radiation to the total absorbed luminosity, 3) $A_{FUV}$, the attenuation in the $FUV$ band.

From the attenuated and unattenuated SED provided for each galaxy, we can directly calculate $f_{\text{dust}} = L_{\text{dust}}/L_{\text{bol}}$. We can also easily calculate the detailed SED of the radiation absorbed by dust, i.e. $L_\lambda^{\text{abs}}$. From this absorbed SED we calculate the contribution due to UV radiation as
\begin{equation}
\xi_{\text{UV}}
=
\dfrac{\int_{\text{UV}} L_\lambda^{\text{abs}}\,{\text{d}}\lambda}{\int L_\lambda^{\text{abs}}\,{\text{d}}\lambda}
=
\dfrac{1}{L_{\text{dust}}} \int_{\text{UV}} L_\lambda^{\text{abs}}\,{\text{d}}\lambda
\end{equation}
where the integral in the numerator covers the UV domain ($\lambda < 3650\, \AA$). Finally, $A_{FUV}$ is computed by convolving both the attenuated and the unattenuated best fit SED with the \textit{GALEX} $FUV$ filter response curve and computing
\begin{equation} \label{eq:afuv}
A_{FUV} = -2.5\log(L_{FUV}^\mathrm{att} / L_{FUV}^\mathrm{unatt}).
\end{equation}

In the subsequent analysis, we correlate $f_{\text{dust}}$ and $\xi_{\text{UV}}$ to a number of other global physical galaxy parameters, such as stellar mass, SFR, and specific star formation rate (sSFR), which is often used in the context of galaxy evolution. We also re-calibrate the often studied relation between IRX vs. the $A_{FUV}$ relation, where we define IRX as TIR/$FUV = L_\mathrm{dust}/L_{FUV}^\mathrm{att}$. Our calibration is done in a self-consistent framework and provides a method for determining $A_{FUV}$ from empirical, observational data, which are relatively easy to obtain. All of these quantities are directly derived from the MAGPHYS model. MAGPHYS has been thoroughly tested, and we limit ourselves to those parameters that are known to be determined well by the fitting, given a good panchromatic wavelength coverage. However, these parameters can still be model dependent. We come back to this in Sect.~\ref{sec:discussion}.

The uncertainties on the attenuation parameters and on the physical quantities $L_\mathrm{dust}$, $M_\star$, SFR, and sSFR are derived through a Monte Carlo bootstrapping method. For each galaxy, we alter the observed fluxes by choosing a random value from a normal distribution centred on the observed flux and with the flux uncertainties as standard deviation. We then re-run the MAGPHYS fitting and derive all our parameters again from the output. We perform 100 iterations per galaxy, sufficient for a reliable estimate of the uncertainty on our parameters.

\section{Results} \label{sec:results}

\begin{figure*}
\centering
\includegraphics[width=0.85\textwidth]{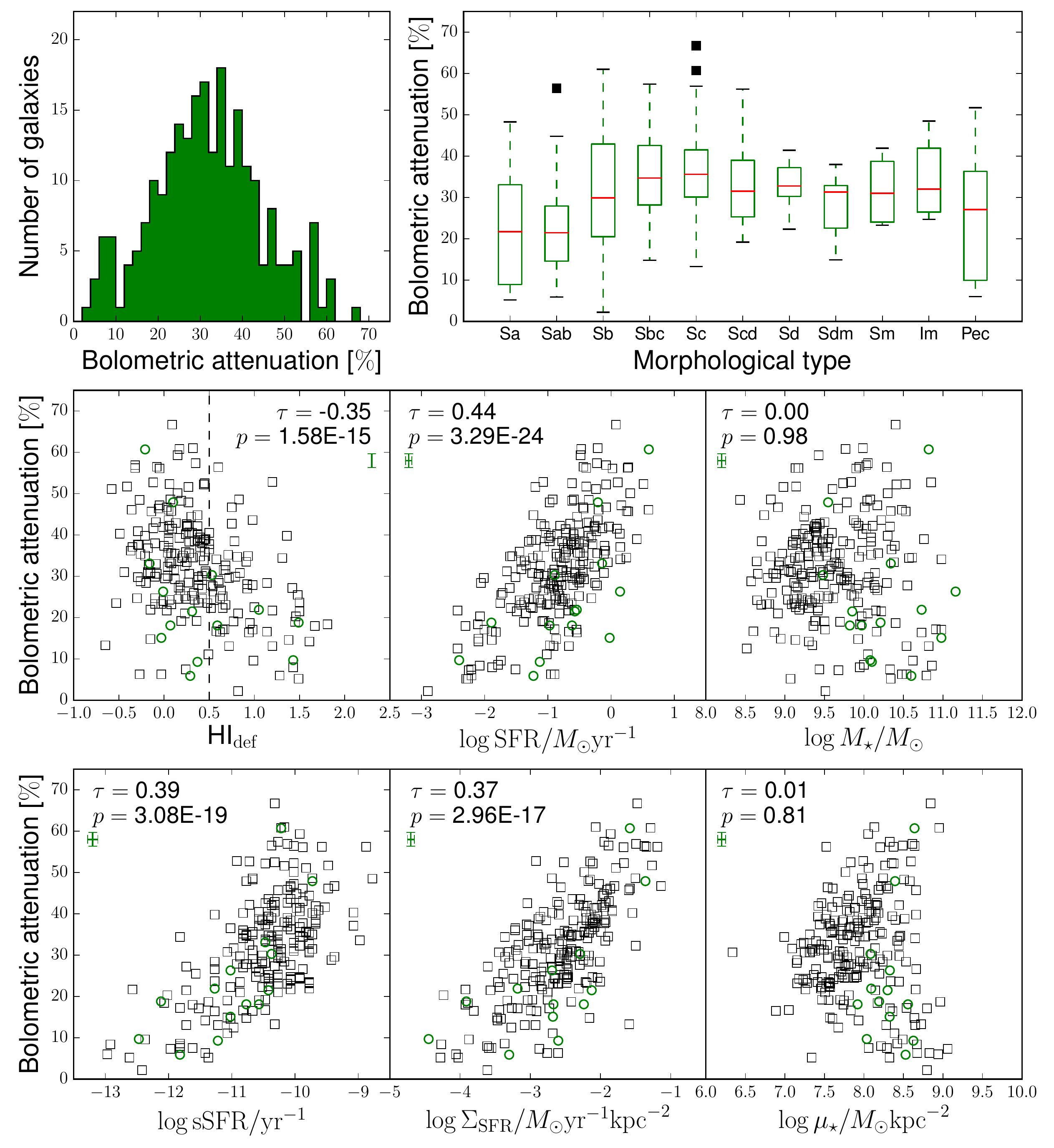}
\caption{Top left: Histogram of the bolometric attenuation $f_{\text{dust}}$ for the galaxies in the sample. Top right: Box plots of $f_{\text{dust}}$ for different morphological types. Red lines indicate the median values, boxes the 1st and 3rd quantiles. Outliers are plotted as black squares. Middle row: Correlations between the bolometric attenuation and HI deficiency, SFR, and stellar mass. Bottom row: Correlations between the bolometric attenuation and sSFR, SFR surface density, and stellar mass surface density. Green circles are classified as strong AGNs. The Kendall's $\tau$ correlation coefficients and corresponding p-values are indicated for each scatter plot. Average error bars are shown in green.}
\label{BolAttenuation.fig}
\end{figure*}

\subsection{The bolometric attenuation} \label{sec:bolatt}

The top left-hand panel of Figure~{\ref{BolAttenuation.fig}} shows a histogram of the bolometric attenuation for all the galaxies in the sample. The mean value obtained for our sample is $\langle f_{\text{dust}} \rangle = 32\%$, which basically reproduces the value obtained by previous studies \citep{1991AJ....101..354S, 1995A&A...293L..65X, 2002MNRAS.335L..41P, Skibba2011}. \citet{1991AJ....101..354S} used the integrated optical and FIR luminosity density of the 60 $\mu$m-selected IRAS Bright Galaxy Atlas \citep[BGS,][]{1987ApJ...320..238S} to estimate $f_{\text{dust}}$. \citet{1995A&A...293L..65X} and \citet{1996A&A...306...61B} improved on this study by using a sample of nearby spiral galaxies that were detected by IRAS and observed in the UV. To correct for the thermal emission from cool dust beyond the IRAS cut-off at 120 $\mu$m and in the MIR region, they used empirical correction factors based on a small sample of 13 nearby galaxies observed in the submm.

\citet{2002MNRAS.335L..41P} used ISO data out to 170~$\mu$m, so they could actually trace the cool dust component in the individual galaxies. These authors attributed their finding the same value of 30\% as obtained by \citet{1991AJ....101..354S} to two factors that work in opposite directions. On the one hand, the BGS is biased towards FIR-bright galaxies, as already
mentioned, and thus favours galaxies that have more dust than the average local spiral galaxy population. On the other hand, estimating the dust luminosity from only IRAS flux densities misses the bulk of the cold dust in galaxies, which implies that the estimates of the bolometric attenuation in \citet{1991AJ....101..354S} are an underestimate of the true bolometric attenuation. It seems a happy coincidence that these two factors, which work in opposite directions, cancel out exactly. 

More recently, \citet{Skibba2011} have analysed the dust/stellar flux $F_d/F_\star$ for the KINGFISH galaxies. For their subset of $35$ nearby spirals,
they find that $\langle \log(F_d/F_\star) \rangle = -0.35$. This is equivalent to $\langle f_{\text{dust}} \rangle = 31\%$. Their method was purely observational, integrating the UV to submm SED through simple linear interpolation. However, they indicate that their sample was not statistically complete.

That we again recover essentially the same number as these earlier studies is remarkable and requires further exploration. First, one would expect that the ISO-based estimates of the dust luminosity might be an underestimate of the real dust luminosity in spiral galaxies. Indeed, most spiral galaxies still have significant  dust emission beyond 170 $\mu$m, so FIR/submm data beyond this wavelength are required to correctly recover the dust temperature distribution, hence dust mass and luminosity \citep[e.g.][]{2010A&A...518L..89G, 2011A&A...532A..56G}. \citet{Skibba2011} note that the lack of SPIRE data led to the underestimation of the total dust luminosity by $17 \%$. On the other hand, they do retrieve the same average value for $f_{\text{dust}}$. But then again their sample is relatively small and contains objects in different environments than the ones from \citet{2002MNRAS.335L..41P}. Additionally, one could expect the mean value of the bolometric luminosity of our study to be higher than the 30\% value obtained by \citet{2002MNRAS.335L..41P} because their sample contained only Virgo Cluster galaxies. It is well-known that spiral galaxies in the Virgo Cluster are generally deficient in atomic gas \citep{1973MNRAS.165..231D, 1983AJ.....88..881G}, which is considered to be due to ram pressure stripping by the hot cluster gas. Recent observations suggest that the same accounts for interstellar dust: Virgo Cluster galaxies contain on average less interstellar dust than similar galaxies in sparser environments \citep{Boselli2006, Cortese2010a, 2012A&A...540A..52C}. 

To investigate whether this has an effect on the mean bolometric attenuation, we plotted the bolometric attenuation as a function of the HI deficiency in the middle left-hand panel of Fig.~\ref{BolAttenuation.fig}. For more information on the HI data for the HRS and the computation of def$_\mathrm{H\textsc{i}}$, we refer to \citet{Boselli2014}. To quantify the strength of the correlations presented in this paper, we computed Kendall's rank coefficient $\tau$ \citep{Kendall1938,Kendall1990}. This non-parametric correlation coefficient is derived from the number of concordant (C) and discordant (D) pairs upon ranking the data: $\tau = \frac{C-D}{C+D}$. A $\tau$ value of one points towards a perfect correlation, while $\tau = -1$ indicates an anti-correlation. The coefficient is zero when no correlation is present. For each $\tau$, an associated p-value can be computed that reflects the chance that the null hypothesis (no correlation) is true. This formalism allowed us to directly compare the correlations we present in this work and quantify the chances of a genuine link between physical parameters.

The correlation between $f_\mathrm{dust}$ and def$_\mathrm{H\textsc{i}}$ is weak, and we find that Kendall's $\tau = -0.35$. However, the chance of no correlation is small ($p = 1.58 \times 10^{-15}$). Most of the HI deficient galaxies (def$_\mathrm{H\textsc{i}}$ $> 0.5$) do occupy the region where $f_\mathrm{dust} < 30\%$. In fact, for the HI deficient galaxies, the average $\langle f_{\text{dust}} \rangle = 0.25$ with a standard deviation of 0.11. On the other hand, galaxies with def$_\mathrm{H\textsc{i}}$ $< 0.5$ have $\langle f_{\text{dust}} \rangle = 0.34$ with a standard deviation of 0.13. A Kolmogorov-Smirnov (K-S) test showed that the probability of both samples coming from the same distribution is $7.35\times 10^{-8}$. This means the bolometric attenuation properties of HI deficient and non-deficient galaxies are significantly different. The trend could indicate that environment and galaxy interactions play a role in governing the amount of starlight that is attenuated. This is in line with the findings of \citet{Cortese2010a, Boselli2014b} and can be interpreted in two ways. One possibility is that rapid quenching of the star formation occurs. When this happens (which is usually the case in a cluster environment), the UV energy output by new stars drops dramatically. Consequently, less energy is absorbed, and the bolometric attenuation fraction drops. Galaxies with quenched star formation due to interactions are usually deficient in atomic gas as well, hence the observed trend. Alternatively, dust may simple get stripped when a galaxy enters a dense environment. This directly causes a decrease in attenuation fractions.

That star formation affects the bolometric attenuation fraction is clear from the middle panel in Fig.~\ref{BolAttenuation.fig}. We observe a positive correlation between $f_{\text{dust}}$ and SFR, with $\tau = 0.44$ and $p = 3.29\times 10^{-24}$. Still there is quite some scatter on the relation: e.g. for Milky Way type galaxies with a SFR of the order of 1~$M_\odot$~yr$^{-1}$, there are galaxies with bolometric attenuation below 20\% and others with values up to almost 60\%. It has been shown that galaxies with high def$_\mathrm{H\textsc{i}}$ usually have low SFR \citep[see][for a review]{Boselli2006}. This is an intuitive link as gas content and SFR are strongly connected through the Schmidt-Kennicutt law \citep{Schmidt1959, Kennicutt1998}. Because SFR anti-correlates with def$_\mathrm{H\textsc{i}}$ and positively correlates with $f_{\text{dust}}$, the observed trend between $f_{\text{dust}}$ and def$_\mathrm{H\textsc{i}}$ may just be an indirect correlation.

Interestingly, there is  a$98 \%$ chance of no correlation ($\tau = 0.00$) with stellar mass, another key parameter in galaxy evolution (Fig~\ref{BolAttenuation.fig}, middle right panel). There is a known positive trend between SFR and galaxy mass \citep[see e.g.][and references therein]{Speagle2014}. Following the above reasoning, one might expect an indirect correlation between $M_\star$ and $f_{\text{dust}}$. It appears that the indirect links are not strong enough to exhibit a trend between those two quantities. This indicates that galaxy mass is no indication of how much of the starlight is absorbed by dust.
\citet{Wang1996} investigated this indirectly through the correlation between the $F_\mathrm{UV}/F_\mathrm{FIR}$ flux ratio and $\Delta V_{H\textsc{i}}$, the velocity dispersion of $H\textsc{i}$ gas. They interpret $\Delta V_{H\textsc{i}}$ as a proxy for stellar mass and the $F_\mathrm{UV}/F_\mathrm{FIR}$ flux ratio as a tracer of the fraction of light from young stars that is escaping the galaxy. In a way, this is the inverse of the attenuation (which is the light not escaping). Their results suggest we should at least see some trend between $f_{\text{dust}}$ and $M_\star$. However, both relations are difficult to compare, not only because different parameters are used, but also because the sample of \citet{Wang1996} is quite different. They investigated nearby galaxies with IRAS detection, while the HRS is K-band-selected. This way, early-type spirals are under-represented in their sample. These objects are usually more massive and lower in attenuation, which would contradict the trend they observe.

We investigated the changes in bolometric attenuation for different morphological types. \citet{2002MNRAS.335L..41P} claimed a strong dependence of the bolometric attenuation on morphological type. For early-type spirals, they found a mean bolometric attenuation of only $15\%$, whereas this rose to $30\%$ for late-type spirals. The top right-hand panel of Figure~{\ref{BolAttenuation.fig}} shows the distribution of the bolometric attenuation as a function of the morphological type of the galaxies in our sample. The average value of $f_{\text{dust}}$ does increase from $23\%$ for Sa galaxies to $37\%$ for Sc galaxies and then remains more or less flat. For each individual class of galaxies, however, there is a significant spread in the values. Even for the Sa class, for example, the values vary between $5\%$ and $50\%$. In fact, for each morphological class, the mean value for $f_{\text{dust}}$ is compatible with the global mean value of $32\%$ within one standard deviation (see Table~{\ref{MeanValues.tab}}). 

The variation is particularly high for galaxies classified as peculiar (Pec). While their median bolometric attenuation fraction is roughly consistent with late-type spirals, this subsample ranges between $5\%$\ and $50 \%$ in $f_{\text{dust}}$. Naively, one would expect higher values, since galaxy mergers are usually associated with peculiar morphologies. Mergers can trigger star formation, which in turn boosts dust production, and can cause more attenuation. We visually inspected these galaxies and found that they cover a wide variety in shape and colour. Some have disk and spiral arms, and others are dust lane elliptical galaxies. None of them, however, show ongoing interaction with another galaxy. Their NUV-r colours range from 2.1 to 5.8, which illustrates the diversity in this class. We therefore do not consider them when investigating trends with morphology.

We ran a K-S test to investigate whether the $f_{\text{dust}}$ values for Sa-Sab galaxies, on one hand, and Scd-Sd, on the other, stem from the same distribution. The chance that both samples are the same is $7.50\times 10^{-4}$. We can thus state a slight increase in bolometric attenuation going from early-type spirals to late types. This is consistent with the previous claims by \citet{2002MNRAS.335L..41P}. The conclusion is that it is impossible to make a sensible estimate of the bolometric attenuation for individual galaxies based on their morphological type. Even for statistical studies, this cannot be done without introducing large errors.

Global extensive parameters, such as stellar mass and SFR, can be prone to scaling effects. A galaxy may contain much dust or form quite a few stars, but if the galaxy is a large and extended object, this will not translate into a strong attenuation. On the other hand, intensive parameters (such as sSFR = SFR/ $M_\star$ and surface density quantities) do hold information about the internal proportions of dust and starlight. One can naively expect galaxies with high sSFR to have more SFR per stellar mass, hence more dust. These objects are denser, which naturally induces more attenuation by dust.

With this motivation, we have plotted the bolometric attenuation as a function of parameters related to the internal galaxy environment in the bottom row of Fig.~\ref{BolAttenuation.fig}. The sSFR is provided by MAGPHYS, and we computed the stellar mass surface density $\mu_\star$ and the SFR surface density $\Sigma_\mathrm{SFR}$ using the $i$ band half-light radius following \citet{Cortese2012a}. There is an upward trend between $f_\mathrm{dust}$ and the sSFR, with $\tau = 0.39$ and $p =1.58\times 10^{-15}$. The correlation with $\Sigma_\mathrm{SFR}$ is about as strong ($\tau = 0.37$ and $p =2.96 \times 10^{-17}$) as with sSFR, but weaker than with SFR. We find no correlation with stellar mass surface density ($\tau = 0.01$ and $p = 0.81$).

The galaxies classified as strong AGNs by Gavazzi et al (in prep.) do not exhibit different values for $f_\mathrm{dust}$. They occupy the same ranges in all of the correlations of Fig.~\ref{BolAttenuation.fig}. This suggests that the central activity does not contribute significantly to the attenuation properties of the galaxy as a whole. For this reason, we did not treat them as a separate subsample when computing the correlation coefficients or mean attenuation values. 

From our analysis with a statistically representative sample of the local universe, we can now confirm that one third of the produced starlight is attenuated by dust. For the bolometric attenuation, we find no tight correlation with main galaxy evolutionary parameters. The strongest trend is an increasing bolometric attenuation with increasing SFR, which proves that more energy is absorbed in the more active galaxies.

\begin{table}
\centering
\caption{Mean value and the standard deviation of the bolometric attenuation $f_{\text{dust}}$ and the UV heating fraction $\xi_{\text{UV}}$ as a function of the morphological type.}
\begin{tabular}{ccr@{$\,\pm\,$}lr@{$\,\pm\,$}l}
\hline\hline\\[-0.5ex]
type & $N_{\text{gal}}$ & \multicolumn{2}{c}{$f_{\text{dust}}$} & \multicolumn{2}{c}{$\xi_{\text{UV}}$} \\
     & & mean & std & mean & std \\
\hline \\
Sa              & 23  & 0.23 & 0.14 & 0.38 & 0.24 \\
Sab             & 20  & 0.23 & 0.12 & 0.36 & 0.19 \\
Sb              & 48  & 0.32 & 0.15 & 0.50 & 0.21 \\
Sbc             & 31  & 0.36 & 0.10 & 0.56 & 0.15 \\
Sc              & 37  & 0.37 & 0.11 & 0.61 & 0.17 \\
Scd             & 29  & 0.34 & 0.10 & 0.67 & 0.16 \\
Sd              & 18  & 0.33 & 0.06 & 0.70 & 0.19 \\
Sdm             & 10  & 0.28 & 0.08 & 0.71 & 0.14 \\
Sm              & 4   & 0.32 & 0.08 & 0.85 & 0.04 \\
Im              & 8   & 0.35 & 0.09 & 0.74 & 0.13 \\
Pec             & 10  & 0.26 & 0.15 & 0.53 & 0.29 \\
total   & 238 & 0.32 & 0.13 & 0.56 & 0.22 \\
\hline
\end{tabular}
\label{MeanValues.tab}
\end{table}

\subsection{The heating of dust in spiral galaxies}

\begin{figure*}
\centering
\includegraphics[width=0.85\textwidth]{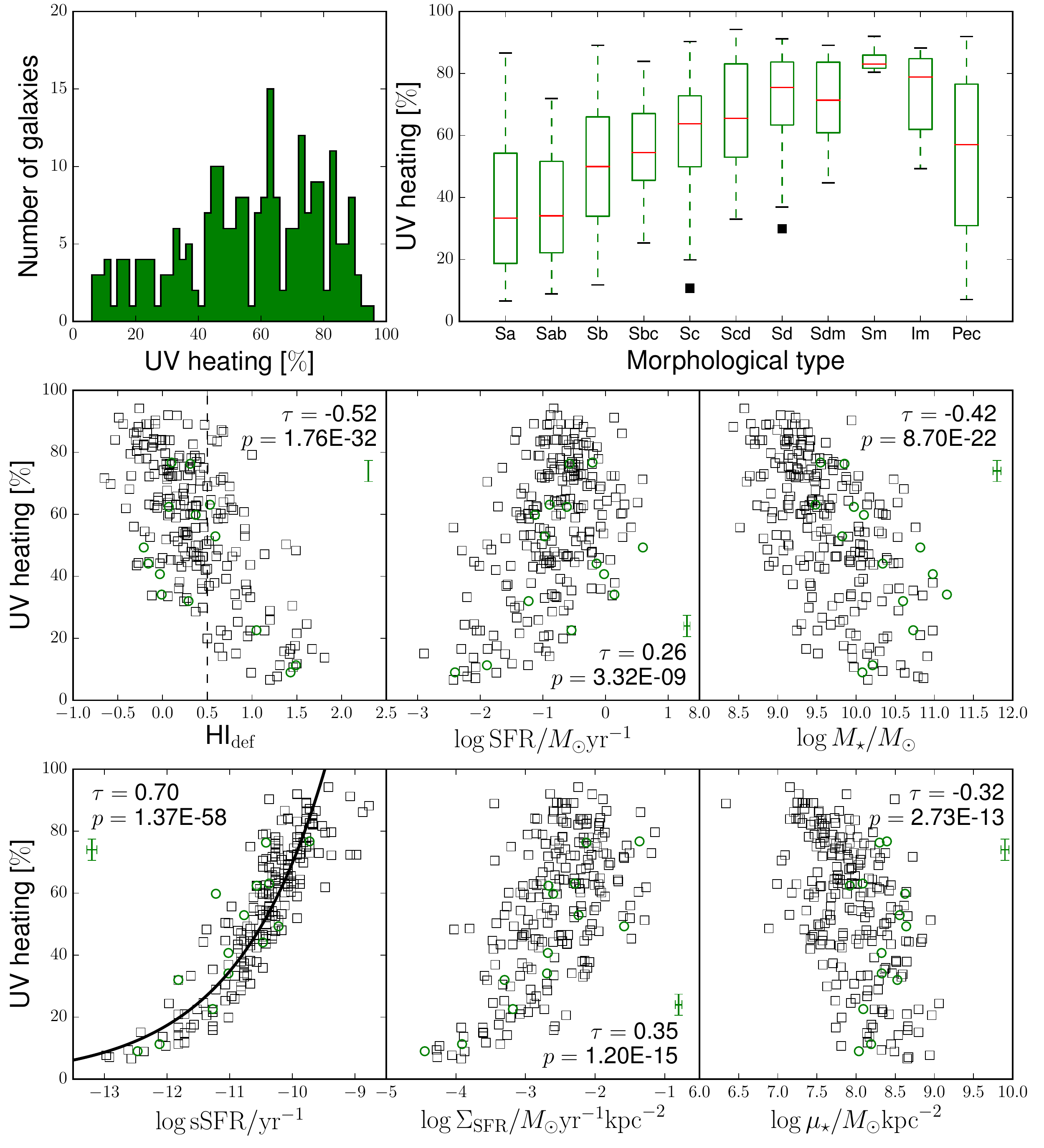}
\caption{Top left: Histogram of the UV heating fraction $\xi_{\text{UV}}$ for the galaxies in the sample. Top right: Box plots of $\xi_{\text{UV}}$ for different morphological types. Red lines indicate the median values, boxes the 1st and 3rd quantiles. Outliers are plotted as black squares. Middle row: Correlations between the UV heating and HI deficiency, SFR, and stellar mass. Bottom row: correlations between the UV heating and sSFR and between the SFR surface density and stellar mass surface density. Green circles are classified as strong AGNs. The Kendall $\tau$ correlation coefficients and corresponding p-values are indicated for each scatter plot. Average error bars are shown in green.}
\label{UVheating.fig}
\end{figure*}

The corresponding results concerning the UV heating fraction (i.e. the percentage of dust luminosity that comes from absorbed UV photons) are shown in Figure~{\ref{UVheating.fig}}. The top left-hand panel shows the histogram of $\xi_{\text{UV}}$ for the 221 galaxies in the sample. The histogram shows a broad distribution that is peaked around $65\%$, but it is skewed towards lower values, such that the average value is only $56\%$. Notably, the full range of values between 0 and 100\% is covered: in some galaxies nearly all of the luminosity absorbed by the dust is optical and NIR radiation, whereas in other galaxies, the dust heating is powered almost exclusively by UV radiation. The distribution of $\xi_{\text{UV}}$ over the different morphological types (top right panel) now shows a clear trend, with the average value increasing from less than $40\%$ for the earliest type spirals to $85\%$ for the Sm galaxies. Within every morphological class, however, there is no uniformity, and a broad range of values is found. We performed the same K-S test as for $f_{\text{dust}}$, i.e. between the combined Sa-Sab sample and the Scd-Sd sample. The probability that both samples are statistically equivalent is $1.15\times 10^{-5}$ for the $\xi_{\text{UV}}$ parameter. Potentially, the trend with morphology may be caused by the fact that early-type spirals have larger bulges and so contain more evolved stars relative to the ongoing star formation. The evolved stars then act as a second heating mechanism in the galaxy, causing the UV heating fraction to go down.

The middle left-hand panel of Fig.~\ref{UVheating.fig} shows that this time there is an anti-correlation with HI deficiency. The trend is stronger than with $f_\mathrm{dust}$ ($\tau = -0.52$ and $p =1.76\times 10^{-32}$), albeit still with some scatter. We find $\langle \xi_{\text{UV}} \rangle = 0.40$ with a standard deviation of 0.23 for HI deficient galaxies (def$_\mathrm{H\textsc{i}}$ $> 0.5$) and $\langle \xi_{\text{UV}} \rangle = 0.63$ with an rms of 0.19 for galaxies with def$_\mathrm{H\textsc{i}}$ $< 0.5$. A K-S test showed that the change of both samples coming from the same distribution is $4.02\times 10^{-8}$. It appears that the environment has a stronger effect on the UV heating fraction than on the bolometric attenuation. We can again interpret this trend in the context of star formation quenching as described in section~\ref{sec:bolatt} or as an effect of dust stripping. This process shuts down an important source of UV light, causing the UV contribution to the total attenuation to go down. This is in line with a significant (but weaker than with $f_\mathrm{dust}$) positive trend with SFR ($\tau = 0.26$ and $p = 3.32\times 10^{-9}$).

In the middle right-hand panel in Fig.~\ref{UVheating.fig}, an inverse trend between $\xi_{\text{UV}}$ and stellar mass can be noted with $\tau = -0.42$ and $p =8.70\times 10^{-22}$. UV radiation dominates the absorbed luminosity in the less massive galaxies, and this portion decreases gradually if we move to higher stellar masses. The picture does not really improve when plotting the surface density parameters $\Sigma_\mathrm{SFR}$ and $\mu_\star$ in the bottom row (middle and right panels). In the correlation with the UV heating fraction and $\mu_\star$, the trend is steep and we find $\tau = -0.32$ with $p =2.73\times 10^{-13}$. That would mean the stellar density has very little influence on the attenuation in the UV. Considering that $\mu_\star$ indirectly traces the radiation field of the older stellar populations, this is not surprising. Old stars in general do not dominate the UV radiation field. The trend between $\Sigma_\mathrm{SFR}$ and $\xi_\mathrm{UV}$ is stronger ($\tau = 0.35$ and $p = 1.20 \times 10^{-15}$) than the one with SFR. When more stars are formed per unit of area, the UV radiation field is higher and more UV light can be absorbed.

Most interesting is the correlation between $\xi_{\text{UV}}$ and the sSFR, as shown in the bottom left-hand panel. Here we obtain a tight correlation, with $\tau = 0.70$ and a very low $p = 1.37\times 10^{-58}$. Galaxies with a low sSFR (${\text{sSFR}}<10^{-11}$~yr$^{-1}$) all have low UV heating, whereas in all galaxies with ${\text{sSFR}}\gtrsim10^{-10}$~yr$,^{-1}$ the absorbed dust luminosity is dominated by UV radiation. Since the scatter on this trend is much smaller compared to the other trends, sSFR seems like the best option for tracing the UV heating in spiral galaxies. A similar trend was also identified by \citet{DeLooze2014}. They used radiative transfer simulations to quantify the dust heating fraction due to young stars and found that high heating fractions correspond with high levels of sSFR and vice versa. We quantified our relation by fitting a power law function to the data points and find
\begin{equation} \label{eq:xiUV_sSFR}
\xi_{\text{UV}} = 7.3^{+1.7}_{-1.4} \times 10^4 \, \mathrm{sSFR}^{0.3017 \pm 0.0086},
\end{equation}
which gives $\xi_{\text{UV}}$ in percentage points for a sSFR in $\mathrm{yr}^{-1}$. Uncertainties on the coefficients were derived through a Monte Carlo method with 1000 iterations.

Both parameters are related to the excess in UV emission to the emission of the evolved stellar population. The more stars that are formed per unit of stellar mass, the stronger the radiation field. This will heat the dust  (mainly through absorption of UV photons) and produce more MIR and FIR dust emission. On the other hand, we do not find a strong correlation with SFR. This is consistent with the fact that galaxies in our sample with higher SFR are simply larger (higher $M_\star$) galaxies. The intrinsic properties of the radiation field are less influenced by this scaling effect. In this case, it is thus more valuable to look at sSFR than at extensive parameters, such as SFR or $M_\star$.

We again note that the AGNs do not stick out as having different $\xi_{\text{UV}}$ values. They follow the trends of the other galaxies in Fig.~\ref{UVheating.fig}. In our subsequent analysis, we no longer plot them as a separate subsample.

\subsection{Calibrating the IRX vs. $A_{FUV}$ relation}

\begin{figure}
\centering
\includegraphics[width=0.45\textwidth]{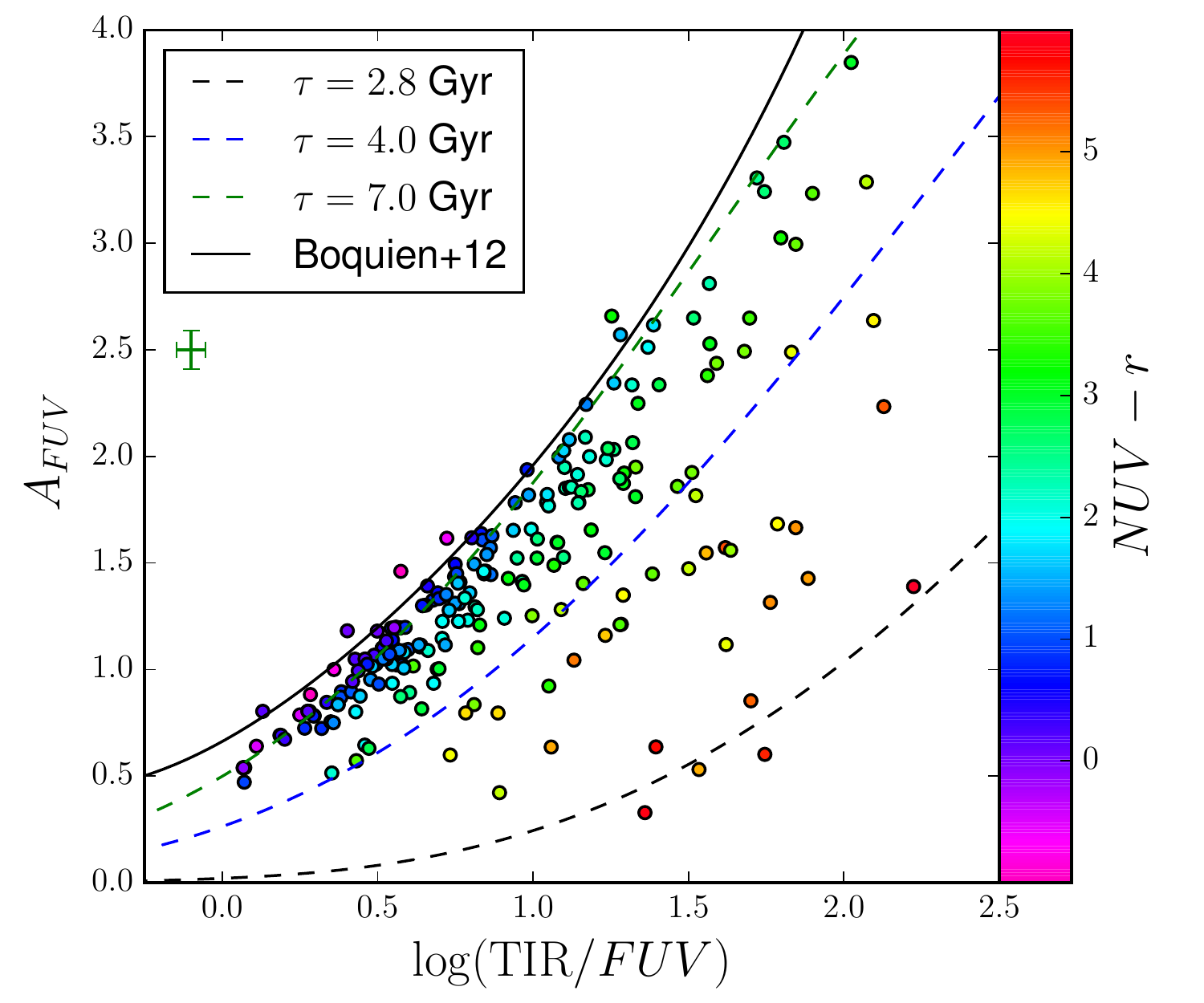}
\caption{$A_{FUV}$ vs. $\log(\mathrm{TIR}/FUV)$ relation for the HRS late-type galaxies, colour-coded according to $NUV-r$ colour. For comparison, a few relations from the calibration by C08 are plotted as dashed lines. Average error bars are shown in green.} 
\label{fig:TIRFUV_AFUV}
\end{figure}

Our models allow us to revisit the IRX vs. $A_{FUV}$ relation for late-type galaxies. There is a strong trend between those two properties, but also a wide spread (see Fig.~\ref{fig:TIRFUV_AFUV}). The main difficulty for constructing this relation is a reliable measurement of $A_{FUV}$. This quantity requires information of the intrinsic, unattenuated SED, which has, up to now, not been easy to obtain. Therefore, previous calibrations of the IRX vs. $A_{FUV}$ relation started from the well-studied and observable IRX vs. $\beta$ relation, where $\beta$ is the slope of the UV SED \citep{Calzetti1994, Meurer1999, Kong2004, Cortese2008, Hao2011}. They then rely on theoretical or semi-empirical considerations to derive $A_{FUV}$. More recently, \citet{Boquien2012} have demonstrated that it is possible to use panchromatic SED fitting as a way to determine $A_{FUV}$ independently and link it to the IRX - $\beta$ relation.

\citet{Cortese2008}, hereafter C08, find that the spread on the IRX-$A_{FUV}$ relation was closely related to the star formation history. They assumed a SFH `a la Sandage' in the formalism of \citet{Gavazzi2002}, which is parametrized by $\tau$, the time after formation at which the star formation in the galaxy peaks. This parameter is related to the sSFR, which is easier to derive from observations. To provide a more empirical calibration of the scatter in the TIR/$FUV$ vs. $A_{FUV}$ relation, we make use of the close correlation of $NUV-r$ and sSFR. The former is easy to obtain, even without making any model asumptions. In our model, we define IRX $\equiv$ TIR/$FUV = L_\mathrm{dust}/L_{FUV}^\mathrm{att}$.  And $A_{FUV}$ was derived using Eq.~\ref{eq:afuv}, which is a direct determination, without having to work through the IRX-$\beta$ relation.

We plot the correlation between IRX and $A_{FUV}$ in Fig.~\ref{fig:TIRFUV_AFUV} and colour code them according to $NUV-r$ colour. A clear, nonlinear but monotonic trend is indeed visible. Galaxies with a high $FUV$ attenuation also have a high TIR/$FUV$ ratio. The scatter in the plane correlates well with $NUV-r$, with blue galaxies exhibiting more $FUV$ attenuation than redder galaxies. The trends make sense in the picture where more active galaxies are bluer and hold relatively more dust (hence more attenuation). This dust is heated to higher temperatures because star formation acts as a second dust heating source. Consequently, the IRX ratio increases. Additionally, we plot the relations found by C08 for $\tau = 2.8$, 4.0, and 7.0, corresponding to galaxies of old, intermediate, and young stellar populations. The theoretical relations by C08 follow the trend closely. The relation for galaxies with old (young) stellar populations follows the data points with high (low) NUV-r colours. This is again comforting for our model, but also opens the opportunity to calibrate this correlation. Finally, in Fig.\ref{fig:TIRFUV_AFUV}, the relation found by \citet{Boquien2012} is shown. This relation was derived from seven local face-on spirals, and they found a strong similarity ($-0.3 < \Delta A_{FUV} < 0.16$) between their relation and previous parametrisations of the IRX-$A_{FUV}$ relation from different samples \citep{Burgarella2005,Buat2011,Hao2011}. These relations are consistent with the blue HRS galaxies ($NUV-r \approx 0$), but do not follow the IRX-$A_{FUV}$ trend for redder galaxies. This again calls for a parametrisation depending on $NUV-r$ or sSFR.

\begin{figure*}
\centering
\includegraphics[width=0.9\textwidth]{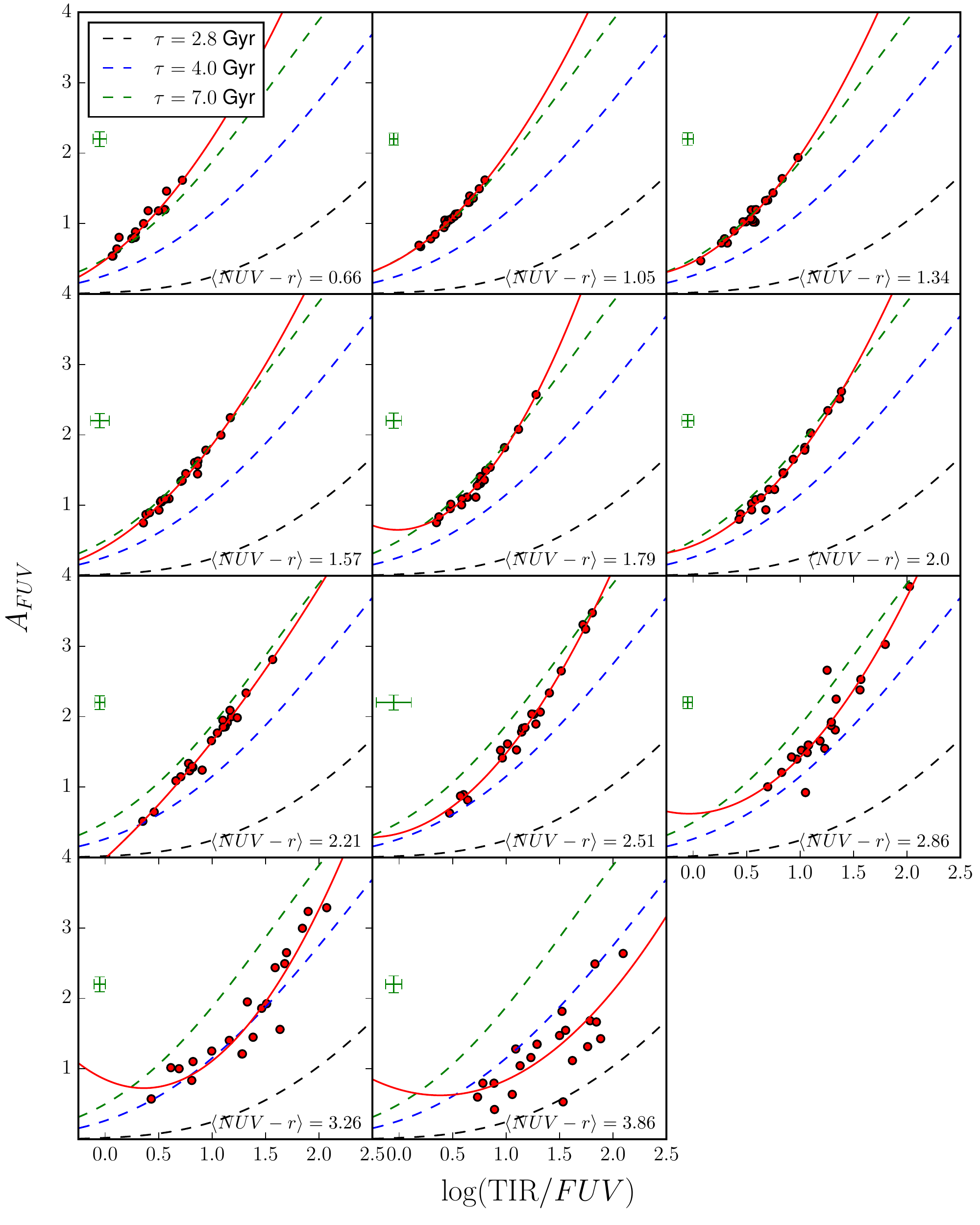}
\caption{$A_{FUV}$ vs. $\log(\mathrm{TIR}/FUV)$ relation separated in different bins of $NUV-r$ colour. The points in each bin are fitted with a second-degree polynomial (solid line). For comparison, a few relations from the calibration by C08 are plotted as dashed lines. Average error bars per bin are shown in green.} 
\label{fig:calibrations}
\end{figure*}

We divided our sample in 11 bins of 20 galaxies according to their $NUV-r$ colour. We chose to fit a second-order polynomial to the data in each bin. The number of data points did not allow us to go to higher order polynomials. Unfortunately, adding more galaxies per bin would increase the spread on $NUV-r$ within one bin. To fit with more general functions (with more free parameters), we need to increase the number of galaxies per bin without increasing the spread on $NUV-r$. This is currently not possible with our dataset. However, we consider this a good trade-off, which covers a wide range in $NUV-r$ with sufficient resolution to be applicable on other sets of late-type galaxies. Figure~\ref{fig:calibrations} shows the best fit polynomial for each sSFR bin, together with a few relations from C08. The best fit parameters are listed in Table~\ref{Calibration.tab}.

\begin{table}
\centering
\caption{Conversion relations for the $\log(\mathrm{TIR}/FUV)$ vs. $A_{FUV}$ relation for different bins of $NUV-r$. The last column shows $\langle \Delta A_{FUV} \rangle$, the average absolute deviation in $A_{FUV}$ from the polynomial relation.}
\begin{tabular}{cccccc}
\hline\hline\\[-0.5ex]
\multicolumn{2}{c}{$\langle NUV-r \rangle$} & \multicolumn{4}{l}{$A_{FUV} = a_1 + a_2x + a_3x^2, x = \log(\mathrm{TIR}/FUV)$} \\
 mean & std & $a_1$ & $a_2$ & $a_3$ & $\langle \Delta A_{FUV} \rangle$ \\
\hline \\
0.66 & 0.25 &  0.48309 & 1.12325 & 0.60186 & 0.05 \\
1.05 & 0.11 &  0.49580 & 0.86097 & 0.63454 & 0.02 \\
1.34 & 0.06 &  0.45683 & 0.77105 & 0.73777 & 0.04 \\
1.57 & 0.08 &  0.41163 & 0.88936 & 0.55688 & 0.04 \\
1.79 & 0.06 &  0.65207 & 0.03586 & 1.13833 & 0.04 \\
2.00 & 0.07 &  0.42749 & 0.58636 & 0.71669 & 0.05 \\
2.21 & 0.07 & -0.01291 & 1.39637 & 0.26219 & 0.05 \\
2.51 & 0.11 &  0.34217 & 0.40083 & 0.73603 & 0.06 \\
2.86 & 0.09 &  0.62276 & 0.05598 & 0.74223 & 0.14 \\
3.26 & 0.12 &  0.84988 &-0.68556 & 0.94567 & 0.21 \\
3.86 & 0.21 &  0.70715 &-0.43529 & 0.56733 & 0.29 \\
\hline\hline
\end{tabular}
\label{Calibration.tab}
\end{table}

This framework provides a physical relation between IRX, $NUV-r$, and $A_{FUV}$ and a more direct way to estimate the attenuation in the $FUV$ band. Both $NUV-r$ and TIR/$FUV$ are relatively easy to determine for large samples of galaxies \citep[e.g. following][]{Galametz2013,Chang2015}. Our calibration allows computing the intrinsic $FUV$ radiation and gives insight into the dust attenuation in a galaxy. On the other hand, if $A_{FUV}$ can be derived (following our approach or any other) together with the $NUV-r$, our relations give a rather precise value for the total infrared luminosity. This quantity is important for studying galaxies near and far, but it is only available for a part of the galaxies in the observed universe.

We do not claim that these polynomial relations are physically motivated. They were used to give a good functional representation of the observed relations. In this respect, they hold within the parameter range of the calibrating data and can be used to interpolate within this range. We caution against extrapolating beyond the calibrated parameter ranges or applying these relations on objects with highly different properties than our calibration sample.

\section{Robustness of the results} \label{sec:discussion}

Our results were derived by a physically motivated model for a galaxy's SED. In the first place, our goal was to investigate whether there are observable trends and which ones are worth pursuing. In that aspect, MAGPHYS provides self-consistent results that can be used to probe these correlations. However, the absolute values of the best fit parameters may differ from those derived from single-band tracers or alternative methods. It is worth investigating model dependencies that can influence our results.

\begin{figure}
\centering
\includegraphics[width=0.5\textwidth]{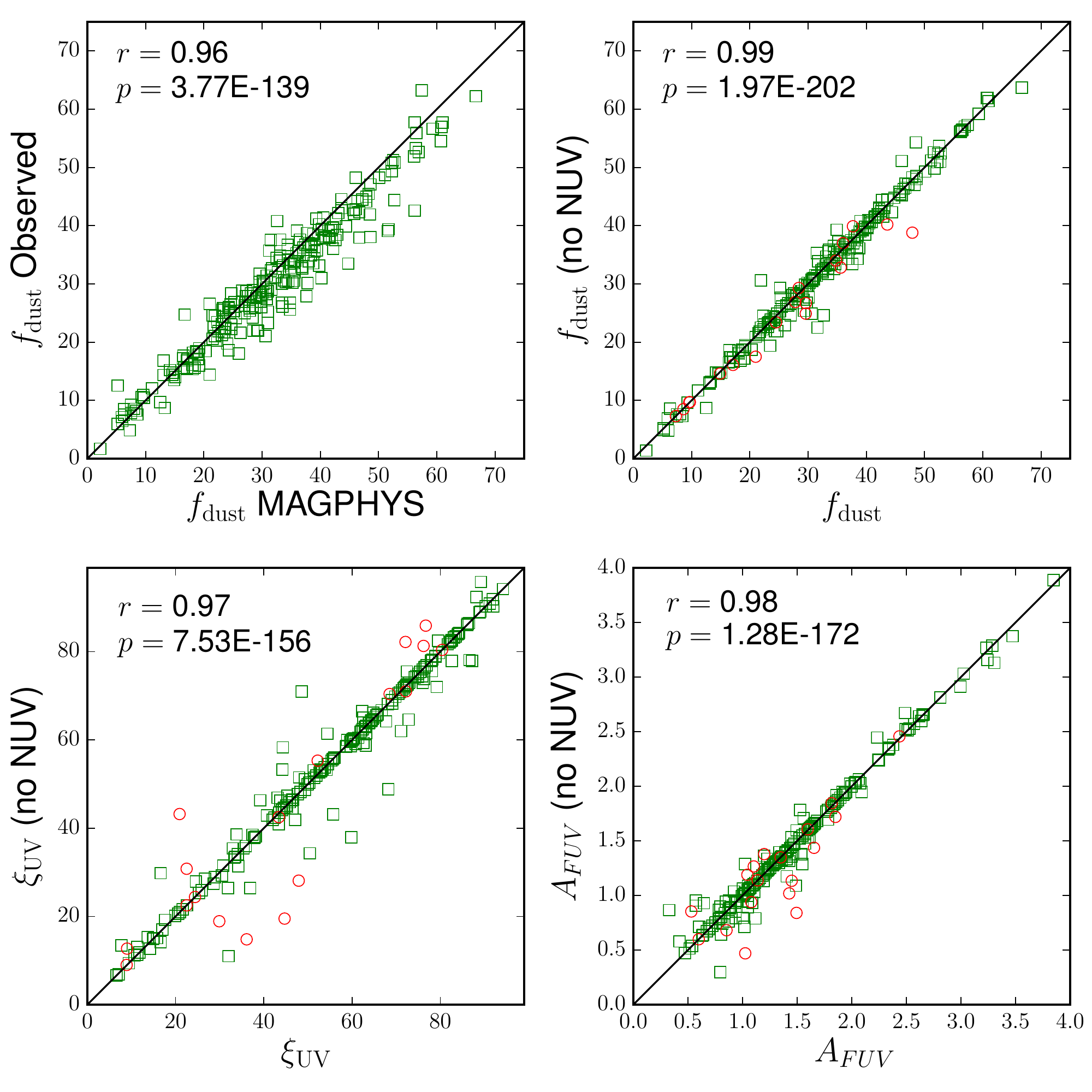}
\caption{Checks to trace potential systematic errors in our method. Top left: $f_\mathrm{dust} \approx L_\mathrm{IR}/ L_\mathrm{bol}$ as derived from integrating the observed SED plotted against $f_\mathrm{dust}$ derived from our fitted model SEDs. Top right: $f_\mathrm{dust}$ as derived from the MAGPHYS fits with (ordinate) and without the $NUV$ data point. Red circles indicate galaxies with no $UV$ data after removing the $NUV$ band. Bottom: Results for $\xi_\mathrm{UV}$ (left) and $A_{FUV}$ (right) of the same run without the $NUV$ flux.} 
\label{fig:checks}
\end{figure}

As a first test, we derived the bolometric attenuation from pure observable quantities without any underlying model assumptions. To do this, we did a linear interpolation (in log-space) between the $FUV$-500 $\mu$m data points. The purely empirical SED was then integrated from the UV to the submm to obtain $L_\mathrm{bol}$. Similarly, we integrated the purely empirical SED from 4 $\mu$m to the submm data points to compute $L_\mathrm{IR}$. The bolometric attenuation can then be estimated as $f_\mathrm{dust} \approx L_\mathrm{IR}/ L_\mathrm{bol}$. The top left-hand panel in Fig.~\ref{fig:checks} shows a very tight correlation between the empirical and model $f_\mathrm{dust}$. The Pearson correlation coefficient $r$ (aimed at probing linear relations) is close to unity, and the chance $p$ of no linear correlation is virtually zero. However, the trend does not follow a strict 1:1 relation; MAGPHYS produces slightly higher $f_\mathrm{dust}$ fractions.

The underlying reason for this situation is that the $L_\mathrm{dust}$ from MAGPHYS is systematically higher than the observed $L_\mathrm{IR}$. The bolometric luminosities do match closely. This discrepancy is most likely linked to the integration interval. For the observed $L_\mathrm{IR}$, we integrated from 4 $\mu$m to the last available submm data point (500 $\mu$m in most cases). This is less straightforward in the case of MAGPHYS, where the dust SED is defined from $0.1-1000$ $\mu$m. Although the SED is virtually zero in the $0.1-3$ micron regime, some flux from hot dust can contribute to the total $L_\mathrm{dust}$ that is not captured in our $L_\mathrm{IR}$. Additionally, PAH peaks are included in the MAGPHYS model, but not in the observed SED owing to a lack of spectral resolution. As a result it is not surprising that MAGPHYS produces slightly higher total dust luminosities. This is also one of the reasons for choosing for a full SED model, rather than an ad hoc integration of the observed SED. The correlations we examined in the previous section still hold when using the empirical $f_\mathrm{dust}$ because the link with the MAGPHYS $f_\mathrm{dust}$ is tight and linear.

In a second test, we verified the validity of the MAGPHYS attenuation model of \citet{Charlot2000}. Interstellar and circumstellar dust extinction are described by a few parameters, and their relative contribution is also a free parameter \citep[for more details, see][]{daCunha2008}. This parametrization creates enough diversity in the attenuation curves to cover a realistic range of attenuation levels. The dust model in MAGPHYS is a rather ad hoc, but physically motivated dust emission model. The dust SED is a combination of modified black bodies with different temperatures and emissivities, and a template for the PAH features based on M17 \citep{daCunha2008}. It is difficult to compare this to self-consistent, physical dust models from e.g. \citet{Zubko2004}, \citet{Draine2007} or \citet{Jones2013}. Those models naturally yield extinction properties that (together with the dust distribution) lie at the basis of the attenuation curve. In MAGPHYS, the extinction model and the dust emission model are in fact two different entities. A handful of dust-related free parameters determine the shape of the MIR-submm SED, but it is the forced coupling with the absorbed starlight that drives the total dust luminosity. This so-called energy balance was modelled so that the total dust luminosity falls within $15\%$ of the total absorbed stellar energy. The attenuation curve is consequently shaped by this combination of stellar and dust SEDs, not by imposing an attenuation or extinction law beforehand. For example, for galaxies with the same optical SED, the galaxy with the highest dust luminosity will have higher values of attenuation.

The reliability of MAGPHYS for different intrinsic galaxy attenuation curves has been studied by \citet{Hayward2015}. They performed MAGPHYS fits of synthetic galaxies created by a combination of radiative transfer and hydrodynamical simulations, where the intrinsic SED is known. They found that MAGPHYS was able to reproduce the intrinsic SED quite well if the attenuation curve did not deviate significantly from the true attenuation curve. In their tests, MW and LMC curves returned reliable fits, while SMC-type attenuation curves showed more discrepancy. It is not possible to measure the true attenuation curve for our sample, but we can argue that these local, star forming galaxies are not particularly low in metallicity \citep{Hughes2013} and fall in the same class of large spirals like the MW. It is therefore likely that they will not have an SMC-like attenuation curve and our models provide a good estimate of the attenuation curve. It is difficult to quantify these model uncertainties, but this falls beyond the scope of this work.

The MAGPHYS extinction model does not include a bump at $NUV$ wavelengths, as observed for the Milky Way or the LMC \citep{Gordon2003, Fitzpatrick2007}. The absence or presence of an $NUV$ bump will affect the model $NUV$ flux. To check any influence on the determination of our attenuation parameters, we re-ran the fitting excluding the $NUV$ data point from the SED of all galaxies. From the new set of models, we again computed $f_\mathrm{dust}$, $\xi_\mathrm{UV}$, and $A_{FUV}$. These $NUV$-less quantities are compared to their original values in Fig.~\ref{fig:checks}. The bolometric attenuation is relatively unaffected by the exclusion of the $NUV$ point. We find a Pearson coefficient $r$ of 0.99. This is expected because the additional extinction by the $NUV$-bump is relatively small compared to the total absorbed energy. The effect is greater for the UV heating fraction, although MAGPHYS still retrieves the same best fit model in most cases and $r = 0.97$. For 18 galaxies, no UV data points are left after removing the $NUV$ point. This leaves only little constraint on the UV attenuation curve, making comparison difficult. Consequently they make up half of the scatter in the plots in the bottom row of Fig.~\ref{fig:checks}. The determination of $A_{FUV}$ appears to be quite robust against the exclusion of the $NUV$ point, especially for higher values of attenuation. At the low attenuation side, MAGPHYS is more likely to find a different best fit model. However, these models still yield an attenuation that is relatively close to the old one, and no outliers are found. This results in a comforting $r$ value of 0.98.

A final concern is whether MAGPHYS is able to capture the star formation history (SFH) of the galaxies in our sample.  MAGPHYS adopts an exponentially declining SFH with a random chance of starbursts occurring over the lifetime of the galaxy. The reliability of this formalism has been tested by \citet{Smith2015} for the same set of hydrodynamical simulations as \citet{Hayward2015} in the discussion above. Although they are able to retrieve the intrinsic SFHs for normal galaxies, it requires marginalising over the library of fitted SEDs. In fact, they find that the SFH of the best fit model is not reliable. Unfortunately, they do not test the influence of the SFH discrepancy on any of the MAGPHYS output parameters, such as stellar mass and SFR. Additionally, for more complex SFHs such as strong bursts and mergers, MAGPHYS is not able to retrieve the SFH even when marginalising over the fitted SEDs. In contrast, when star formation is quenched in a galaxy, a truncated SFH model is necessary according to Ciesla et al. (submitted). They fitted their truncated SFH model to the same sample as ours (HRS late-type galaxies) using CIGALE \citep{Noll2009}. They found little difference with fitting results of a double exponentially declining SFH with a short burst for normal spirals. However, HI deficient spirals were usually better fitted by the truncated SFH, with the main difference being a better fitting UV part. However, the SFH they compare with differs from the MAGPHYS SFH, which uses an exponential model with random short bursts.  Without this addition of multiple short bursts, it appears to be difficult to produce an UV SED that fits both \textit{GALEX} data points. In our MAGPHYS fits, even for the HI deficient galaxies, the SED matches the \textit{GALEX} observations for virtually all galaxies.

From these two recent studies, it is not clear how strong the influence of the SFH is on the key parameters we use in this work. MAGPHYS has been tested and proved to give consistent estimates of main galaxy parameters for various samples of local star forming galaxies \citep[e.g.][and reference therin]{daCunha2008,daCunha2010,Smith2012,Driver2015}. Although the SFH can influence parameters, such as stellar mass and SFR during the evolution of a galaxy, there is a strong degeneracy, and many SFH models can reproduce the same observed $M_\star$ and SFR. 

\citet{Michalowski2014} investigated the effect of the SFH parametrisation on stellar mass for a set of simulated submm galaxies. They found that MAGPHYS slightly overestimates the true stellar masses. We can compare the SFRs from this work with the ones derived in \citet{Boselli2015}. They determined the SFR for HRS late-type galaxies using several observational tracers and also using SED fitting. We find strong correspondence in SFR across the sample, with Pearson's correlation coefficients over 0.85. There is, however, a systematic offset of about 0.1-0.3 dex between our values (which are lower) and the different tracers from \citet{Boselli2015}. Without going into detail, we attribute these systematic offsets to differences in the modelling. This includes different assumptions in the SFH and has also been found by \citet{Pacifici2015} for a sample of galaxies at redshifts 0.7-2.8. On the other hand, \citet{Buat2014} investigated the SFR for a sample of $z>1$ galaxies using SED fitting
again with different SFHs. They found that the SED-derived SFRs are slightly \textit{higher} than the classical IR$+FUV$ tracer. The bottom line here is that while this may influence the absolute location of the correlations in the plots with $f_\mathrm{dust}$ and $\xi_\mathrm{UV}$, it does not neutralise the trends we found.

It is even more difficult to quantify how the SFH influences the attenuation curve and the derived parameters $f_\mathrm{dust}$, $\xi_\mathrm{UV}$, and $A_{FUV}$. There is a strong need for more research in this area. In particular, the mutual influence of the dust model and the SFH model on retrieving the correct attenuation curve of galaxies is an important question to address in the future.

\section{Conclusions} \label{sec:conclusions}

We have performed panchromatic (UV-submm) SED modelling of the HRS late-type galaxies. Our main goal was to quantify the total amount of absorbed energy by dust. We used the power of MAGPHYS to investigate the difference between the observed and the intrinsic (dust free) SEDs of our sample and derive three key parameters; the bolometric dust fraction $f_\mathrm{dust}$, the UV heating fraction $\xi_\mathrm{UV}$, and the $FUV$ attenuation $A_{FUV}$. We connected these quantities to key parameters of galaxy evolution, such as $M_\star$, SFR, and sSFR. Our main conclusions are:\\
\begin{itemize}
  \item The mean bolometric attenuation  $\langle f_{\text{dust}} \rangle$ is $32\%$ for our sample. For local, star forming galaxies about one third of the energy produced by stars is absorbed by dust. This number is, quite surprisingly, in line with previous estimates derived from small and incomplete samples.
  \item We confirmed a weak trend between $f_{\text{dust}}$ and morphological type. A broad range of $f_{\text{dust}}$ values were found for each type of galaxy. The strongest correlation ($\tau = 0.44$) was found with SFR, but still with significant scatter.
  \item The mean UV heating fraction $\langle \xi_{\text{UV}} \rangle = 0.56$, but with a broad distribution. For our sample, more than half of the stellar energy is absorbed in the UV domain.
  \item There is a clear trend between $\xi_\mathrm{UV}$ and morphological type with significantly higher UV heating in galaxies of later type. We also found a strong correlation with sSFR, a known indicator of the heating of dust. The relation is not linear, and we provided a power law fit in equation \ref{eq:xiUV_sSFR}.
  \item We find no evidence that the presence of a strong AGN in a galaxy affects the attenuation properties that galaxy.
  \item We revisited the IRX vs. $A_{FUV}$ relation with both quantities derived directly from the best fit models, offering a self-consistent framework. We calibrated this relation for different values of $NUV-r$, using a second-order polynomial. The best fitting relations are given in Table~\ref{Calibration.tab}. This framework allowed us to estimate $A_{FUV}$ based on quantities that are relatively easy to obtain and to determine the FIR properties of galaxies lacking observations in this regime.  
\end{itemize}
The HRS is designed to provide a concise view of the large galaxies in our local universe. The results of this work should therefore be representative for nearby spirals. The derived relations can be applied to a larger set of local galaxies and can be compared with a similar analysis at higher redshift. The latter in particular could yield important insight into the evolution of energy absorption and reprocessing by dust.

MAGPHYS and other energy balance codes are ideal tools for performing this kind of study in a relatively fast and straightforward way. However, the results are still an interpretation through the underlying galaxy model inherent to the code. While the results appear to be robust, it is still uncertain how well the star formation history and the attenuation curve are reproduced. The effect of a discrepancy in one or both of these two key elements on our parameters is difficult to quantify and requires a more dedicated investigation.

\begin{acknowledgements}
S.V, M.B., and I.D.L. gratefully acknowledge the support of the Flemish Fund for Scientific Research (FWO-Vlaanderen). M.B. acknowledges financial support from the Belgian Science Policy Office (BELSPO) through the PRODEX project "\textit{Herschel}-PACS Guaranteed Time and Open Time Programs: Science Exploitation" (C90370). \\
The authors wish to thank G. Gavazzi and J.F. Otegi for providing AGN classifications for the HRS galaxies. \\
This work was done within the CHARM framework (Contemporary physical challenges in Heliospheric and AstRophysical Models), a phase VII Interuniversity Attraction Pole (IAP) programme organised by BELSPO, the BELgian federal Science Policy Office. \\
We thank all the people involved in the construction and the launch of \textit{Herschel}. SPIRE was developed by a consortium of institutes led by Cardiff University (UK) including Univ. Lethbridge (Canada); NAOC (China); CEA, LAM (France); IFSI, Univ. Padua (Italy); IAC (Spain); Stockholm Observatory (Sweden); Imperial College London, RAL, UCL-MSSL, UKATC, Univ. Sussex (UK); and Caltech, JPL, NHSC, Univ. Colorado (USA). This development has been supported by national funding agencies: CSA (Canada); NAOC (China); CEA, CNES, CNRS (France); ASI (Italy); MCINN (Spain); SNSB (Sweden); STFC and UKSA (UK); and NASA (USA). HIPE is a joint development (are joint developments) by the \textit{Herschel} Science Ground Segment Consortium, consisting of ESA, the NASA \textit{Herschel} Science Center, and the HIFI, PACS and SPIRE consortia.\\
\end{acknowledgements}

\bibliographystyle{aa} 
\bibliography{allreferences}

\end{document}